\documentclass[12pt]{article}
\usepackage{amsmath,
graphicx, enumerate}
\usepackage[letterpaper,top=0.5in, bottom=1in, left=0.8in, right=0.8in, headsep=0.in, centering]{geometry}
\usepackage{amssymb}
\usepackage{epsfig}
\usepackage[normalem]{ulem}

\usepackage{soul}

\title
{The de Broglie Wave as a Localized Excitation\\
 of the Action Function
}

\author{Gregory I. Sivashinsky\\  
Sackler Faculty of Exact Sciences,
School of Mathematical Sciences, \\
Tel Aviv University, Tel Aviv 69978, Israel}
\date{}
\begin{document}

\maketitle
\begin{center}
{\bf Abstract} 
\end{center}
\noindent The Hamilton-Jacobi equation of relativistic quantum mechanics is revisited.  The equation 
is shown to permit
 solutions in the form of breathers (nondispersive oscillating/spinning solitons), displaying simultaneous particle-like and wave-like behavior adaptable to the properties of the de Broglie clock. Within this formalism the de Broglie wave acquires the meaning of a localized excitation of the classical action function.\\
\noindent The  problem of quantization in terms of the breathing action function is discussed. 

\bigskip

\noindent {\bf PACS}: 03.65-w, ~03.65Pm, ~ 03.65.-b

\medskip

\noindent {\bf  Key words}:   de Broglie waves; breathers and solitons; wave-particle duality

\setcounter{equation}{0}

\newpage
\noindent{\bf 1. Introduction}\\
\noindent  In John Bell's book ``Speakable and Unspeakable in Quantum Mechanics'', he asks the question ``What is it that `waves' in wave mechanics?  In the case of water waves it is the surface of the water that waves.  With sound waves the pressure of the air oscillates ... In the case of the waves of wave mechanics we have no idea what is waving ...'' [1].\\
\noindent  In this paper we suggest examining the possibility that it is the Hamilton's action function that waves, an approach which seems not to have been explored previously.\\ 
\noindent Consider the relativistic Hamilton-Jacobi (HJ) equation for a particle in an electromagnetic field,

\begin{eqnarray}\label{1}
\left(1/c^2\right)\left(\partial S/\partial t+e U\right)^2-\left(\nabla S-e {\mathbf A}/c\right)^2=m^2c^2
\end{eqnarray}

\noindent Here $U , {\mathbf A}$ are scalar and vector potentials of the field obeying the Lorentz calibration condition,
\begin{eqnarray}\label{2}
\left(1/c\right) \partial U/ \partial t  +  {\rm div} {\mathbf A} = 0
\end{eqnarray}

\noindent  The trajectory ${\mathbf x}(t)$ of the particle is governed by the equation [2],
\begin{eqnarray}\label{3}
\frac{d {\mathbf x}}{dt} = - \frac{c^2\left(\nabla S^{(0)}-e {\mathbf A}/c\right)}{\partial S^{(0)}/\partial t+e U} \quad ,
\end{eqnarray}

\noindent where $S^{(0)}$ is an appropriate action function of the system.
\noindent  As  is well known, the trajectories ${\mathbf x}(t)$ are characteristics of Eq.\:(1) [2,3], which are, in turn, the traces of small perturbations.  Indeed, let's represent the action function as
\begin{eqnarray}\label{4}
S=S^{(0)}+s
\end{eqnarray}

\noindent where $s$ is a perturbation.  Assuming $s$ to be small, Eq.\:(1) yields the linear equation,
\begin{eqnarray}\label{5}
\left(1/c^{2}\right) \left(\partial S^{(0)}/ \partial t + e U \right) \partial s / \partial t - \left(\nabla S^{(0)}-e {\mathbf A} /c \right) \nabla s = 0 \quad ,
\end{eqnarray}

\noindent whose characteristics are  identical to those of Eq.\:(1).\\
\noindent Let
\begin{eqnarray}\label{6}
{\mathbf f} \left({\mathbf x}, t \right) = {\mathbf c} 
\end{eqnarray}

\noindent be a set of independent integrals of Eq.\:(3), $\mathbf c$ being a constant vector.  Then the general solution of Eq.\:(5) may be written as,
\begin{eqnarray}\label{7}
s=s\left[{\mathbf f} \left({\mathbf x}, t \right) \right]
\end{eqnarray}

\noindent  The perturbation $s$ is therefore advected along the trajectory ${\mathbf x} (t)$ with the velocity ${\mathbf v}=d{\mathbf x}/d t$.
\noindent  If the perturbation is localized enough it will mimic the motion of the particle.  
\noindent As an example, consider the case of a free particle $(U=0, {\mathbf A}=0)$, where
\begin{eqnarray}\label{8}
S^{(0)}=-Et+ {\mathbf p} {\mathbf \cdot}{\mathbf x} \qquad \left(E^{2}/c^{2}=p^{2}+m^{2}c^{2}\right)
\end{eqnarray}

\noindent  Eq.\:(5) then yields,
\begin{eqnarray}\label{9}
s=s\left({\mathbf x} - {\mathbf v} t \right),
\end{eqnarray}

\noindent where
\begin{eqnarray}\label{10}
{\mathbf v}=c^2 {\mathbf p}/E
\end{eqnarray}

\noindent  In the linear approximation the perturbation is advected without changing its shape.  However, in the nonlinear description, due to the Huygens principle, the perturbation will gradually decay thereby implying stability (albeit nonlinear) of the regular solution $S^{(0)}$.\\
\noindent  The question is whether it is possible to modify the HJ equation (1) so that the new equation would allow for localized nonspreading and nondecaying perturbations (excitations) of the regular action function.  Moreover, if the localized excitation breathes (oscillates/spins), one would end up with a dynamic model for a particle with quantum-like features.  As we intend to show, this kind of behavior can be successfully reproduced by the conventional quantum Hamilton-Jacobi (QHJ) equation,
\begin{eqnarray}\label{11}
\left(1/c^2\right)\left(\partial S/\partial t+e U\right)^2 - \left(\nabla S- e {\mathbf A} /c\right)^2 = m^2c^2+i \hbar \Box S \quad ,
\end{eqnarray}

\noindent whose capacity, it transpires, has not been fully explored.
\bigskip
\bigskip

\noindent {\bf 2. A free particle} \\
\noindent As  is well known, the QHJ equation (11) is a transformed version of the linear Klein-Gordon (KG) equation
\begin{eqnarray}\label{12}
\left(1/c^2\right) \left(\partial /\partial t + i e U /\hbar \right)^2 \Psi - \left(\nabla - i e {\mathbf A}/c \hbar \right) ^2 \Psi + \left(mc/\hbar \right)^2 \Psi=0,
\end{eqnarray}

\noindent obtained through the substitution,
\begin{eqnarray}\label{13}
\Psi= \exp \left(i S / \hbar \right)
\end{eqnarray}

\noindent  Consider first the case of a free particle  $(U=0,{\mathbf A}= 0)$ where Eq.\:(12) becomes
\begin{eqnarray}\label{14}
\Box \Psi+(mc/\hbar)^2 \Psi=0 \qquad (\Box=(1/c^2) \partial^2/\partial t^2 - \nabla^2)
\end{eqnarray}

\noindent  The KG equation (14) allows for a two-term spherically symmetric solution
\begin{eqnarray}\label{15}
\Psi= \exp [-i(mc^2/\hbar)t]+ \alpha \exp (-i \omega t)j_0(kr)
\end{eqnarray}

\noindent where
\begin{eqnarray}\label{16}
\omega = c \sqrt{k^2+(mc/\hbar)^2},
\end{eqnarray}

\noindent $r=\sqrt{x^2+y^2+z^2}$, $\alpha $ is a free parameter, and
\begin{eqnarray}\label{17}
j_0(kr)= \sin (kr)/kr
\end{eqnarray}

\noindent is the zeroth-order spherical Bessel function.\\

\noindent  In terms of the action function $S$, by virtue of (13), Eq.\:(15) readily yields
\begin{eqnarray}\label{18}
S=-mc^2t-i \hbar \ln \left\{1+ \alpha \exp \left[-i\left( \frac{\omega - mc^2} {\hbar}\right)t\right]j_0(kr)\right\}
\end{eqnarray}

\noindent  Here the first term corresponds to the classical action function,  $S^{(0)}=-mc^2t$, for a free particle in the rest system while the second term represents its localized nondispersive excitation.  
 
 \noindent  Let's set the frequency of oscillations in Eq.\:(18) in accordance with the de Broglie postulate that each particle at rest can be linked to an internal `clock' of frequency $mc^2/\hbar$.
 \noindent  The frequency $\omega$ in Eq.\:(15) should therefore be specified as
 \begin{eqnarray}\label{19}
 \omega=2(mc^2/\hbar)
 \end{eqnarray}
 
 \noindent  Hence, by virtue of Eq.\:(16),
 \begin{eqnarray}\label{20}
 k=\sqrt{3}(mc/\hbar)
 \end{eqnarray}
 
 \noindent Eq.\:(18) thus becomes,
 \begin{eqnarray}\label{21}
 S= - mc^2t-i\hbar \ln \left\{1+ \alpha \exp \left[-i \left(\frac{mc^2}{\hbar}\right) t \right] j_0 \left[\sqrt{3}\left(\frac{mc}{\hbar}\right)r \right]\right\}
 \end{eqnarray}

\bigskip 
\noindent Note that in Eq.\:(21) the frequency is not affected by the nonlinearity of the system, preserving its value irrespective of the breather intensity, $\left|\alpha\right|$.

\noindent Away from the breather's core ($r \gg\hbar/mc$ ) the oscillations become monochromatic,
\begin{eqnarray}\label{22}
S=-mc^2t-i\alpha\hbar\exp\left[-i\left(\frac{mc^2}{\hbar}\right)t\right] j_0 \left[\sqrt{3}\left(\frac{mc}{\hbar} \right) r \right]
\end{eqnarray}

\noindent Similar to oscillations of an ideal pendulum,  the breather (21) is stable to small perturbations. The stability follows from the linearity of the KG equation.  Due to the linearity there is no coupling between the basic solution (15) and its perturbation, which also obeys the KG equation.  Therefore, if the initial perturbation is small it will remain so indefinitely.  

\noindent Until now we have dealt with a particle at rest.  For a particle moving at a constant velocity         $v$     along, say,        $x$  - axis, the corresponding expression for the action function is readily obtained from Eq.\:(21) through the Lorentz transformation,
\begin{eqnarray} \label{23}
t\to \frac{t-xv/c^2}{\sqrt{1-(v/c)^2}}\;, \qquad x\to\frac{ x-vt}{\sqrt{1-(v/c)^2}}
\end{eqnarray}

\noindent The transformed Bessel function      $j_0$          will then mimic the motion of the classical particle while the transformed temporal factor $\exp\left[-i (mc^2/\hbar)t\right]$ will turn into the associated 
de Broglie wave. The modulated de Broglie wave thus acquires the meaning of a localized excitation of the action function displaying simultaneous particle-like and wave-like behavior.\\


\noindent If, as is conventional, we associate the gradients    $-\partial S/\partial t, \nabla S$                            
with the particle energy         $E$        and momentum      ${\bf p}$, then the Einstein relation
$(1/c^2)E^2=p^2+m^2c^2$
appears to hold far from the     $\hbar/mc$ - wide breather's core, or on average over the entire breather.  
The correspondence with classical relativistic mechanics  is therefore complied with. 




\noindent In addition to spherically symmetric breathers, Eq.\:(14) also permits  asymmetric breathers, spinning around  some axis.  In the latter case the second term of Eq.(15) should be replaced by
\begin{eqnarray}\label{24}
\alpha\exp\left[-2i\left(\frac{mc^2}{\hbar}\right)t+in\phi\right]j_l\left[\sqrt{3}\left(\frac{mc}{\hbar}\right)r\right]P_l^n\left(\cos\theta\right),
\end{eqnarray}							
where       $j_l$, $P_l^n$    are high-order spherical Bessel and Legendre functions.  
\newpage

\noindent {\bf 3.  Quantization}\\
\noindent The next question is how to reproduce quantization directly in terms of the breathing  action function.  The geometrically simplest situation, where such an effect readily manifests itself, is the periodic motion of an otherwise free particle over a closed interval   $0<x<d$.  In this case the field-free version of Eq.(11) must be considered jointly with two boundary conditions,
\begin{eqnarray}\label{25}
\partial S(0,y,z,t)/\partial t=\partial S(d,y,z,t)/\partial t \:, \nonumber \\
\partial S(0,y,z,t)/\partial x=\partial S(d,y,z,t)/\partial x
\end{eqnarray}

\noindent Any classical action function for a free particle,
\begin{eqnarray}\label{26}
 S=-Et+px  
\end{eqnarray}

\noindent is clearly a solution of this problem.  However, in the case of a breathing 
action function the situation proves to be different.  Thanks to the boundary conditions (25), the moving breather interacts with itself, and this may well lead to its self-destruction unless some particular conditions are met.                                         

\noindent Consider first the simplest case of a particle at rest $(v=0)$.  The pertinent solution is readily obtained by converting the problem for a finite interval into a problem for an infinite interval ($-\infty<x<\infty$) filled with a      $d$-periodic train of standing breathers, assumed to be spherical for simplicity.  The resulting action function then reads,
\begin{eqnarray}\label{27}
S=-mc^2t-i\hbar\ln\left\{ 1+\alpha\exp\left[-i\left(\frac{mc^2}{\hbar}\right)t\right]\sum_{k}\left( j_0\right)_{d,0}^{(k)}\right\},
\end{eqnarray}

\noindent where 										
\begin{eqnarray}\label{28}
\left(j_0\right)_{d,0}^{(k)}=\frac{\sin \left[\sqrt{3}\left(mc/\hbar\right)r_{d,0}^{(k)}\right]}{\sqrt{3}\left(mc/\hbar\right)r_{d,0}^{(k)}}\;,
\end{eqnarray}
\begin{eqnarray}\label{29}
r_{d,0}^{(k)}=\sqrt{(x-kd)^2+y^2+z^2} \qquad (k=0,\pm 1,\pm 2,...)
\end{eqnarray}
Here the second subscript stands for    $v=0$.     

\noindent The action function for a moving particle $(v\ne0)$  is obtained from (27) (28) (29) through the Lorentz transformation (23), provided  $d$  is replaced by   $d/\sqrt{1-(v/c)^2}$. The latter step is needed to balance the relativistic contraction, and thereby to preserve the spatial period ($d$)  of the system.  The resulting action-function thus becomes, 
\begin{eqnarray}\label{30}
  S=-Et+px-i\hbar\ln \left\{1+\alpha\exp\left[i\left(\frac{-Et+px}{\hbar}\right)\right]\sum_{k}\left(j_0\right)_{d,v}^{(k)}\right\},         
\end{eqnarray}
where
\begin{eqnarray}\label{31}
\left(j_0\right)_{d,v}^{(k)}=\frac{\sin\left[\sqrt{3}\left(mc/\hbar\right)r_{d,v}^{(k)}\right]}{\sqrt{3}\left(mc/\hbar\right)r_{d,v}^{(k)}}\;,
\end{eqnarray}
\begin{eqnarray}\label{32}
r_{d,v}^{(k)}=\sqrt{\left(\frac{x-vt-kd}{\sqrt{1-(v/c)^2}}\right)^2+y^2+z^2}
\end{eqnarray}
The spatial 	$2\pi\hbar/p$ - periodicity of  $\exp\left[i\left(-Et+px\right)/\hbar\right]$								
is compatible with the spatial     $d$   - periodicity of      
$\sum_{k}\left(j_0\right)_{d,v}^{(k)}$
 only if
\begin{eqnarray}\label{33}
dp=2\pi n\hbar \qquad (n=0,1,2,...)\;,
\end{eqnarray}
which recovers the familiar Bohr-Sommerfeld quantum  condition.\\  
\noindent The above solution (30)-(33) may be easily adapted for the problem of a particle rebounding between two perfectly reflecting walls, $x=0$ and $x=d/2$.  To handle the double-valuedness of the pertinent action function the trajectory of the particle, following 
the Einstein topological approach [4], should be placed on the double-sheeted strip, $0<x<d/2$, $-\infty<y<\infty$, $z=\pm0$.  Thereupon the problem reduces to the previous one.

\bigskip

\noindent{\bf 4. A particle in a slowly varying field}\\
\noindent  The de Broglie postulate holds at least for breathers exposed to slowly varying potentials, characterized by spatio-temporal scales much larger than $\hbar/mc, \hbar/mc^2$.
\noindent  Indeed, as may be readily shown, for slowly varying potentials, Eq.\:(21) is modified to

\begin{eqnarray}\label{34}
S=-\left(mc^2+eU\right)t+ \frac{e}{c} \mathbf A \mathbf \cdot \mathbf x - i \hbar \ln \left\{1+ \alpha \exp \left[-i \left( \frac{mc^2}{\hbar} \right) t \right] j_0 \left[\sqrt{3}\left( \frac{mc}{\hbar} \right) r \right]\right\}
\end{eqnarray}

\noindent  So, unlike the action function as a whole, the frequency of its oscillations in the rest system is not affected by the field.  This invariance is untenable for the wave function $\Psi$ (13), which therefore cannot serve as a physically objective representation of the de Broglie clock.\\
\noindent For a particle moving in a slowly varying field the `fast' spatio-temporal coordinates $\mathbf x,t$ in Eq.\:(34) should be subjected to the Lorentz transformation, with the velocity $\mathbf v$ regarded as a slowly varying vector.

\noindent  The action function (34) and its Lorentz transformed version pertain to the interior of the breather (inner solution).  Away from the breather's core the action function is described by the regular breather-free solution of the QJH equation $S^{(0)}$, involving only large spatio-temporal scales (outer solution).  The inner solution is clearly affected by the outer solution through the velocity field $\mathbf v$, while the reverse influence does not take place, at least not for the leading order asymptotics.  The simplest picture emerges in the nonrelativistic semiclassical limit where the uniformly valid approximate solution may be represented as,

\begin{eqnarray}\label{35}
S=-mc^2t+S_{c}-i\hbar \ln \left\{1+ \alpha \exp \left[-i \left(\frac{mc^2}{\hbar}\right)t\right]\exp \left[i \left(\frac{ m \mathbf v \mathbf \cdot\mathbf x}{\hbar} \right)\right]j_0 \left[\sqrt{3}\left(\frac{mc}{\hbar}\right)r \right]\right\}
\end{eqnarray}


\noindent Here $S_c$ is the classical action function governed by the equation
\begin{eqnarray}\label{36}
\frac{\partial S_c}{\partial t} + \frac{1}{2m} \left(\nabla S_c \right)^{2}+eU=0
\end{eqnarray}
The argument $r$ in $j_0\left[\sqrt{3}\left(mc/\hbar \right) r \right]$ is defined as
\begin{eqnarray}\label{37}
r= \left| \mathbf x - \mathbf x_p \left(t \right)\right|\quad ,
\end{eqnarray}
where  $\mathbf x_p\left(t\right)$ is the classical trajectory of the particle. 

\begin{eqnarray}\label{38}
\frac{d \mathbf x _{p}}{dt} = \frac{1}{m} \mathbf p  \qquad \left(\: \mathbf p = m \mathbf v =\nabla S_{c}\: \right )
\end{eqnarray}

\noindent For bound motions the requirement of spatial periodicity of $\exp [ im (\mathbf v \cdot \mathbf x ) /\hbar]$ leads to a modified Bohr-Sommerfeld quantum condition,

\begin{eqnarray}\label{39}
\oint \mathbf p \cdot d \mathbf x = 2\pi n \hbar \qquad \left(n\gg 1\right)
\end{eqnarray}

\noindent where, following Einstein's topological approach [4], the loop integral (39) is evaluated along topologically nontrivial loops on the torus formed by the whole family of appropriate classical trajectories.

\bigskip
\noindent {\bf 5. Concluding remarks}\\
\noindent  The proposed formalism is certainly related to de Broglie's double solution program [2].  Yet, unlike the latter, in the current model the breather is guided by a regular, nonwaving action function $S^{(0)}$ rather than by a guiding $\Psi$-wave solution.  The guiding action function and its localized excitation (breather) are coupled through the nonlinear QHJ equation (11).  
  


\noindent At this stage it is difficult to see whether the amended QHJ formalism is indeed adequate enough to reproduce all the basic features of quantum-mechanical phenomenology.  In any case, the preliminary observations presented here already show that a mathematical representation of unified wave-particle behavior is quite feasible, even within the framework of the conventional QHJ equation.

\newpage

\noindent {\bf Acknowledgments}



\noindent These studies were supported in part by the Bauer-Neumann Chair in Applied Mathematics and Theoretical Mechanics. 

\end{document}